\documentclass[seceq]{ptptex}

\usepackage{graphicx}

\newcommand{\be}{\begin{equation}}
\newcommand{\ee}{\end{equation}}
\newcommand{\ba}{\begin{eqnarray}}
\newcommand{\ea}{\end{eqnarray}}
\newcommand{\ban}{\begin{eqnarray*}}
\newcommand{\ean}{\end{eqnarray*}}

\newcommand{\bef}{\begin{figure}}
\newcommand{\eef}{\end{figure}}
\newcommand{\bce}{\begin{center}}
\newcommand{\ece}{\end{center}}

\title{Jet-induced gauge field instabilities in the quark-gluon plasma
}

\author{
Massimo \textsc{Mannarelli} and Cristina \textsc{Manuel}
}

\inst{Instituto de Ciencias del Espacio (IEEC/CSIC), \\
Campus Universitat Aut\`onoma de Barcelona, Facultat de Ci\`encies, Torre C5 E-08193 Bellaterra (Barcelona), Spain
}

\abst{We discuss the  properties of the collective modes of a system composed by a thermalized quark-gluon plasma traversed by a relativistic jet of partons. The  transport equations obeyed by the components of the plasma and of the  jet are studied  in the Vlasov approximation. Assuming  that the  partons in the jet  can be described with a tsunami-like distribution function we derive the expressions
of the dispersion law of the collective modes.
Then the behavior of the unstable gauge modes of the system is analyzed for various values of the velocity of the jet, of the momentum of the collective modes and of the angle between these two quantities.
We find that  the most unstable modes are those with momentum  orthogonal to the velocity of the jet, and the
effect is stronger for ultrarelativistic jet velocities.
Our results suggest a new possible collective mechanism for the description of the jet quenching phenomena in heavy ion collisions. }

\begin{document}

\maketitle

\section{Introduction}

One of the methods for unveiling the properties of matter produced in ultrarelativistic heavy-ion
collisions is to study the propagation  properties of high $p_T$ partons generated  by hard scatterings in the initial stage of the collision.
When the jet of partons travels across the medium it loses  energy  and degrades, mainly by radiative processes (see \cite{Kovner:2003zj} for reviews). The energy and momentum of the jet are   absorbed   by the plasma and result in an increased  production of  soft hadrons in some given directions.

We  propose a novel mechanism
\cite{Mannarelli:2007gi,Mannarelli:2007hj} for describing how the
jet loses energy and momentum while travelling in a thermally
equilibrated   quark-gluon plasma (QGP). Since the jet of
particles  is  not in thermal equilibrium with the QGP it perturbs
and destabilizes the system  inducing the generation of gauge
fields, even if the jet of colored particles is neutral over a
coarse grain volume. Some of these gauge modes are unstable and
grow  exponentially fast in time absorbing the kinetic energy of
the  jet.

The novel mechanism we propose for the QGP is   well-known in the context of traditional
plasma physics \cite{Kra73}. The study of the interaction of a relativistic stream of particles with a  plasma is a topic of interest in different fields of physics,
 ranging from inertial confinement fusion,
astrophysics and cosmology. When the particles of the stream carry electric charge, plasma instabilities develop,
leading to an initial stage of fast growth of the gauge fields.

The study of chromo instabilities is  a very active field of research nowadays
(see \cite{Mrowczynski:2006ad} for a review), as it has been claimed that they may speed up the isotropization
and equilibration process in an anisotropic system. Here we discuss another physical setting where
plasma instabilities may play also an important role in heavy ion collisions.

In order to study the system composed by the jet and the plasma  we have employed two different methods.
In \cite{Mannarelli:2007gi}    both the plasma and the jet  are described using a fluid approach. This  approach developed in \cite{Manuel:2006hg} has been derived from kinetic theory expanding the transport equations in moments of momenta and truncating the expansion at the second moment level. The system of equations is then closed with an equation of state relating  pressure and energy density.
The fluid  approach  has several advantages with respect to the underlying kinetic theory. The most remarkable one is that  one has to deal with  a set of  equations much simpler than those of kinetic theory. Then one can    easily generalize the fluid equations  to  deal with more complicated systems. This is a strategy that has been successfully followed in the  study of
different dynamical aspects of non-relativistic electromagnetic plasmas \cite{Kra73}, even if one is considering a system far away from its hydrodynamical regime.

In the second approach  we consider  the same setting,  {\it i.e.}  an equilibrated plasma traversed by a relativistic jet, but  we use kinetic theory  instead of the fluid approach. Transport theory provides a well controlled framework for studying the properties of the quark-gluon plasma in the weak coupling regime, $g \ll 1$.
Indeed it is well known that the physics of long distance scales in an equilibrated weakly coupled
QGP can be described within semiclassical transport equations \cite{Braaten:1989mz,Bla93,Kel94}. In this approach the
hard modes,  with typical energy scales of order $T$,  are treated as (quasi-)particles which propagate in the background of the soft modes, whose energies are equal or less than $gT$, which are treated as classical gauge fields.
This program has been very successful for  understanding some dynamical aspects of the soft gauge fields in an almost equilibrated QGP \cite{Blaizot:2001nr,Litim:2001db}.

In this talk we only review the results of the above second approach. Details of the comparison of the two approaches,
that give similar results, can be found in Ref.~\cite{Mannarelli:2007hj}.


\section{Kinetic theory approach}

We consider a system composed by a quark-gluon plasma traversed by a jet of partons. We  assume that  the system is initially in a  colorless and thermally equilibrated  state and we will study the behavior of  small deviation  from  equilibrium. The distribution function
 of  quarks is
\begin{equation}
Q(p,x) = f^{\rm eq.}_{FD}(p_0) + \delta Q(p,x) \,,\,\,\,\,
\end{equation}
where  the various quantities are hermitian matrices in the fundamental representation of $SU(3)$
(we have suppressed color indices) and where
$f^{\rm eq.}_{FD}(p_0) = \frac{1}{e^{p_0/T} + 1}$
is the (colorless) Fermi-Dirac  equilibrium distribution functions.
A similar decomposition can be done for the distribution function of antiquarks $\bar Q(p,x)$, and
for gluons $G(p,x)$ that we will not write down here (see Ref.~\cite{Mannarelli:2007hj} for details).

Also the jet is assumed to be initially  colorless, but
small color fluctuations are present:
\begin{equation}
W_{\rm jet}(p,x) = f_{\rm jet} (p) + \delta W_{\rm jet}(p,x) \,.
\end{equation}
For the initial jet distribution function we will not consider a thermal distribution function. We will indeed approximate the distribution function of the jet with a colorless tsunami-like form \cite{Pisarski:1997cp}
\begin{equation}
\label{tsunami}
f_{\rm jet}(p) =  \bar n \, \bar u^0 \;
\delta^{(3)}\Big({\bf p}
- \Lambda \, \bar {\bf u} \Big) \;,
\end{equation}
that describes a system of particles of   constant density $\bar n$, all moving with the same  velocity $\bar u^\mu = (\bar u^0, \bar {\bf u}) = \gamma (1, {\bf v})$, where $\gamma$ is the Lorentz factor and   $\Lambda$ fixes the scale of the energy of the particles.

This distribution function is adequate for describing a uniform
and sufficiently dilute system of particles, and it represents an
extreme crude model of a jet of energetic partons, chosen to
simplify our analytical estimates. Of course, it is possible to
use more involved distribution functions, assuming that the
density of particles composing the jet is not uniform and there is
a spread in momentum.

The distribution function of quarks satisfy the following transport equation
\begin{eqnarray}
p^{\mu} D_{\mu}Q(p,x) + {g \over 2}\: p^{\mu}
\left\{ F_{\mu \nu}(x), \partial^\nu_p Q(p,x) \right\}
&=& C \;,
\label{transport-eq}
\end{eqnarray}
whereas  the distribution functions of antiquarks, gluons and of the particles of the jet satisfy similar equations. Here $g$ is the QCD coupling constant, with  $\{...,...\}$ we denote the anticommutator, $\partial^\nu_p$ is
the four-momentum derivative, and $D_{\mu}$ is a covariant derivative.
In Eq.(\ref{transport-eq})  $C$ represents the collision term. However,
for time scales shorter than the mean free path time the collision term can be neglected, as typically done in the so-called Vlasov approximation.

The knowledge of the distribution function allows one to compute the associated color current, which in a self-consistent
treatment enters as a source term in the Yang-Mills equation.  It is important to note that at very short time scales
 the different components of the system formed by the plasma and the jet interact with each other only through the generated average gauge fields.

The contribution to the  polarization tensor
of  the particles of species $\alpha$ (where $\alpha$ refers to quarks, antiquarks, gluons or
the partons of the jet) in the Vlasov approximation is
\begin{equation}
\label{Pi-kinetic}
\Pi^{\mu \nu}_{ab, \alpha}(k) = - g^2 C^{\alpha}_F \delta_{ab}\int_p
f_{\alpha} (p) \;
{ (p\cdot k)(k^\mu p^\nu + k^\nu p^\mu) - k^2 p^{\mu} p^{\nu}
- (p\cdot k)^2 g^{\mu\nu} \over(p\cdot k)^2} \;,
\end{equation}
where $a,b$ are color indices and $C^\alpha_F$ is the value of the quadratic Casimir associated with the particle specie $\alpha$ which takes values
$1/2$ and $3$ for the fundamental and adjoint representations of $SU(3)$, respectively.
The momenta measure is defined as
\begin{equation}
\label{measure}
\int_p \cdots \equiv \int \frac{d^4 p}{(2\pi )^3} \:
2 \Theta(p_0) \delta (p^2-m^2_\alpha) \;,
\end{equation}
where $m_\alpha$ is the mass of the particle of specie $\alpha$.
For simplicity we  assume that the particles belonging to the plasma are massless.  Instead the particles of  the jet  have a non-vanishing mass.

When $f_\alpha(p)$ is a thermal equilibrated distribution function, Eq.~(\ref{Pi-kinetic})
reduces to the form of the hard thermal loop (HTL) polarization tensor \cite{Braaten:1989mz,Bla93,Kel94}.  However  for the tsunami-like distribution function, Eq.~(\ref{tsunami}),
the polarization tensor obviously takes a different form.

The gauge fields  obey the Yang-Mills equation
\begin{equation}
\label{yang-mills1}
D_{\mu} F^{\mu \nu}(x) = \delta j^\nu_t (x) = \delta j_{p}^{\nu}(x) + \delta j_{\rm jet}^{\nu }(x)\; ,
\end{equation}
where we have defined
\begin{equation}
\label{col-current}
\delta j^{\mu }_p(x) = -\frac{g}{2} \int_p p^\mu \;
\Big[ \delta Q( p,x) - \delta \bar Q ( p,x)+  2 \tau^a {\rm Tr}\big[T^a \delta G(p,x) \big]\Big] \;,
\end{equation}
which describes the plasma color current, and
\begin{equation}
\label{jetcol-current}
\delta j_{\rm jet}^{\mu }(x) = -\frac{g}{2} \int_p p^\mu \;
 \delta W_{\rm jet}( p,x)  ,
\end{equation}
which describes the fluctuations of the  current associated with the jet.

Equation (\ref{yang-mills1}) together with Eq.~(\ref{transport-eq})
form a set of equations that has to be solved  self-consistently. Indeed the gauge fields which are solutions of the Yang-Mills
equation enter into the  transport equations of every particle species and,  in turn, affect the evolution of the distribution functions.

\section{Collective modes in the system composed by the QGP and jet}
\label{QGP-JET}

We now consider the collective modes of the system composed by  an equilibrated QGP traversed by a jet of particles. We are interested in very short time scales when the Vlasov approximation can be employed. The  effect of the beam of particles is to induce a color current, which provides a contribution to the polarization tensor. The polarization tensor of the whole system is additive in this short time regime, meaning that
\be
\Pi^{\mu \nu}_{t}(k) = \Pi^{\mu \nu}_{p}(k) + \Pi^{\mu \nu}_{\rm jet}(k) \, ,
\ee
where $\Pi^{\mu \nu}_{p}(k)$ and  $\Pi^{\mu \nu}_{\rm jet}(k)$ are the polarization tensor of the plasma and of the jet respectively. The total dielectric tensor is given by
\be
\label{total-dielectric}
 \varepsilon^{ij}_{\rm t}(\omega,{\bf k}) = \delta^{ij} + \frac{\Pi^{ij}_t}{\omega^2}  \,,
\ee
and the dispersion laws of the collective modes of the whole system  can  be determined solving  the equation
\be
\label{dispersion-T}
 {\rm det}\Big[ {\bf k}^2 \delta^{ij} -k^i  k^j
- \omega^2 \varepsilon^{ij}_{\rm t}(k)  \Big]  = 0 \,.
\ee

The solutions of this equation depend on ${|\bf k|}$, ${|\bf v|}$,
$\cos\theta={\bf \hat k \cdot \hat v}$,
$\omega_t^2 = \omega^2_{\rm p}+\omega^2_{\rm jet}$
where $\omega_{p}$ and $\omega_{\rm jet}$ are the plasma frequency of the QGP and of the jet, respectively,
and on
\be
b = \frac{\omega^2_{\rm jet}}{\omega^2_{p} +\omega^2_{\rm jet} } \ ,
\ee

We find $ \omega^2_{\rm jet} = g^2 \bar n/ 2 \Lambda$, so it depends on the density of the jet as well
as the mean energy that their partons carry.

 Clearly, when the plasma and the jet do not interact, they have stable collective modes. However, once we consider the composed system of plasma and jet interacting via mean gauge field interactions, unstable gauge modes appear.

\begin{figure}[h]
\begin{center}
\includegraphics[width=14pc]{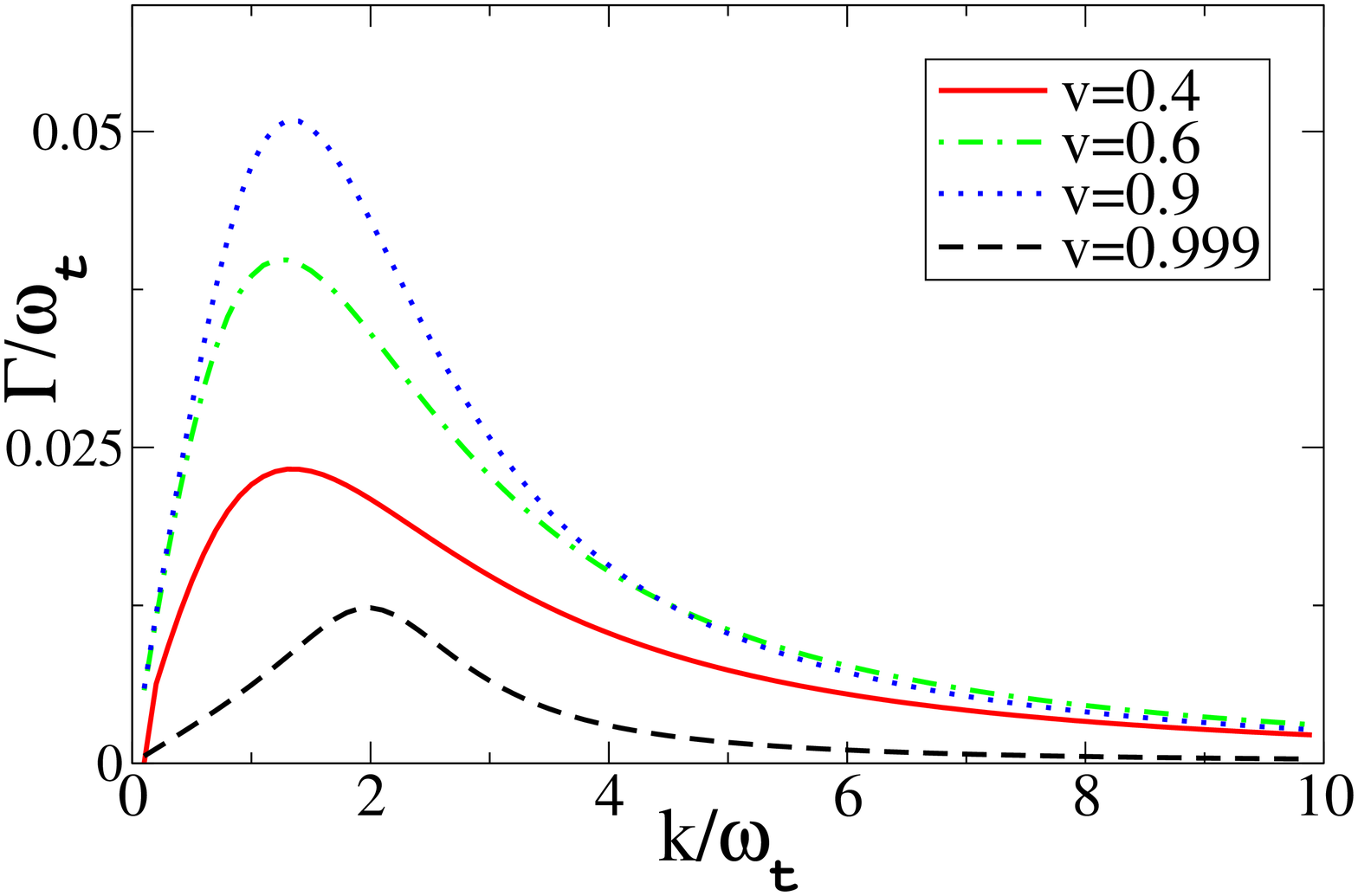}
\includegraphics[width=14pc]{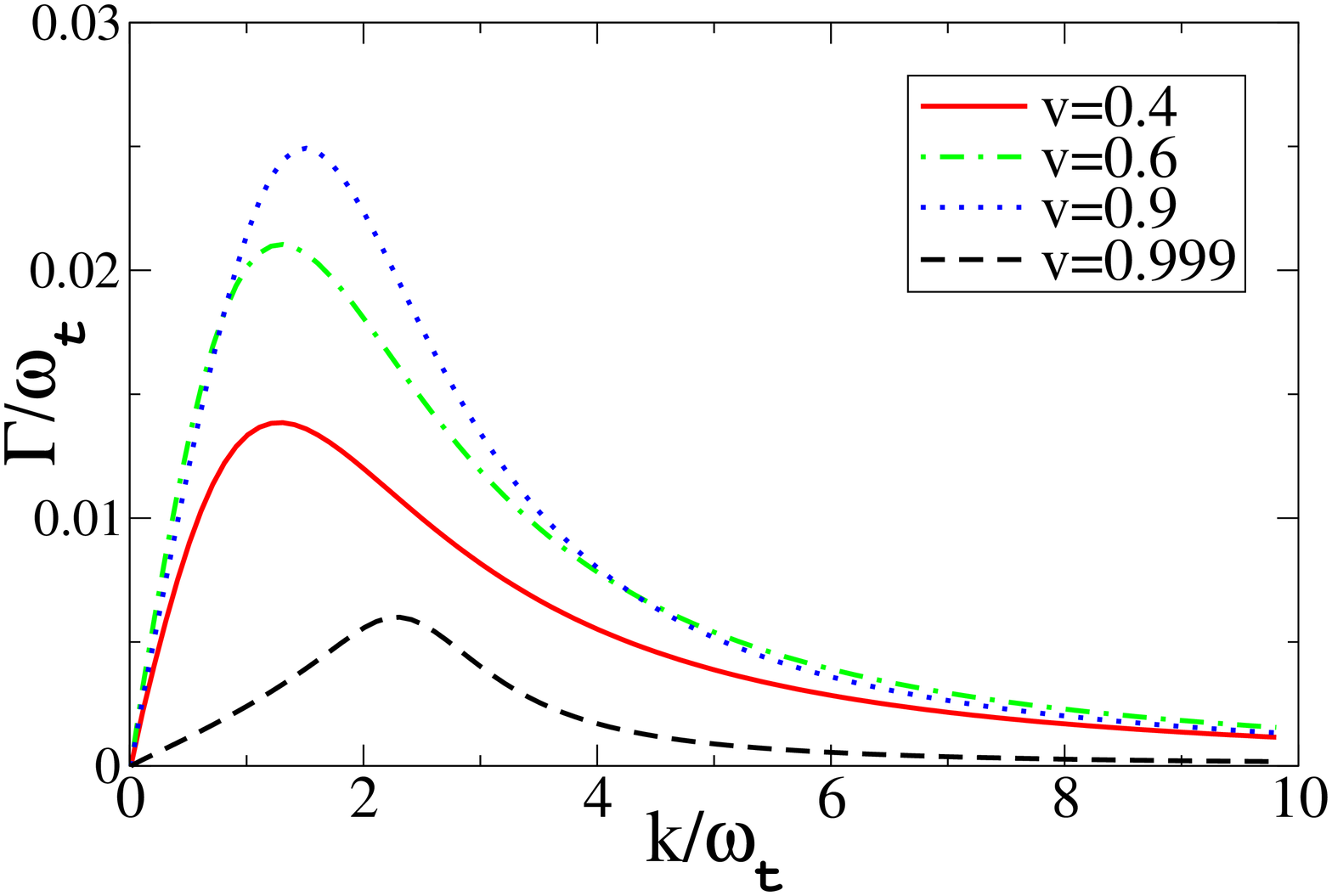}
\caption{\label{para} Imaginary part of the dispersion law of the unstable longitudinal mode for the system composed by a plasma and a jet in the case $\bf k \parallel v$ as a function of the momentum of the mode at  $b=0.1$  (left) and at $b=0.02$ (right) for four different values of the velocity of the jet,  $|{\bf v}|$. \vspace{1cm}} \label{Parafig1}
\end{center}
\end{figure}

\begin{figure}[h]
\begin{center}
\includegraphics[width=14pc]{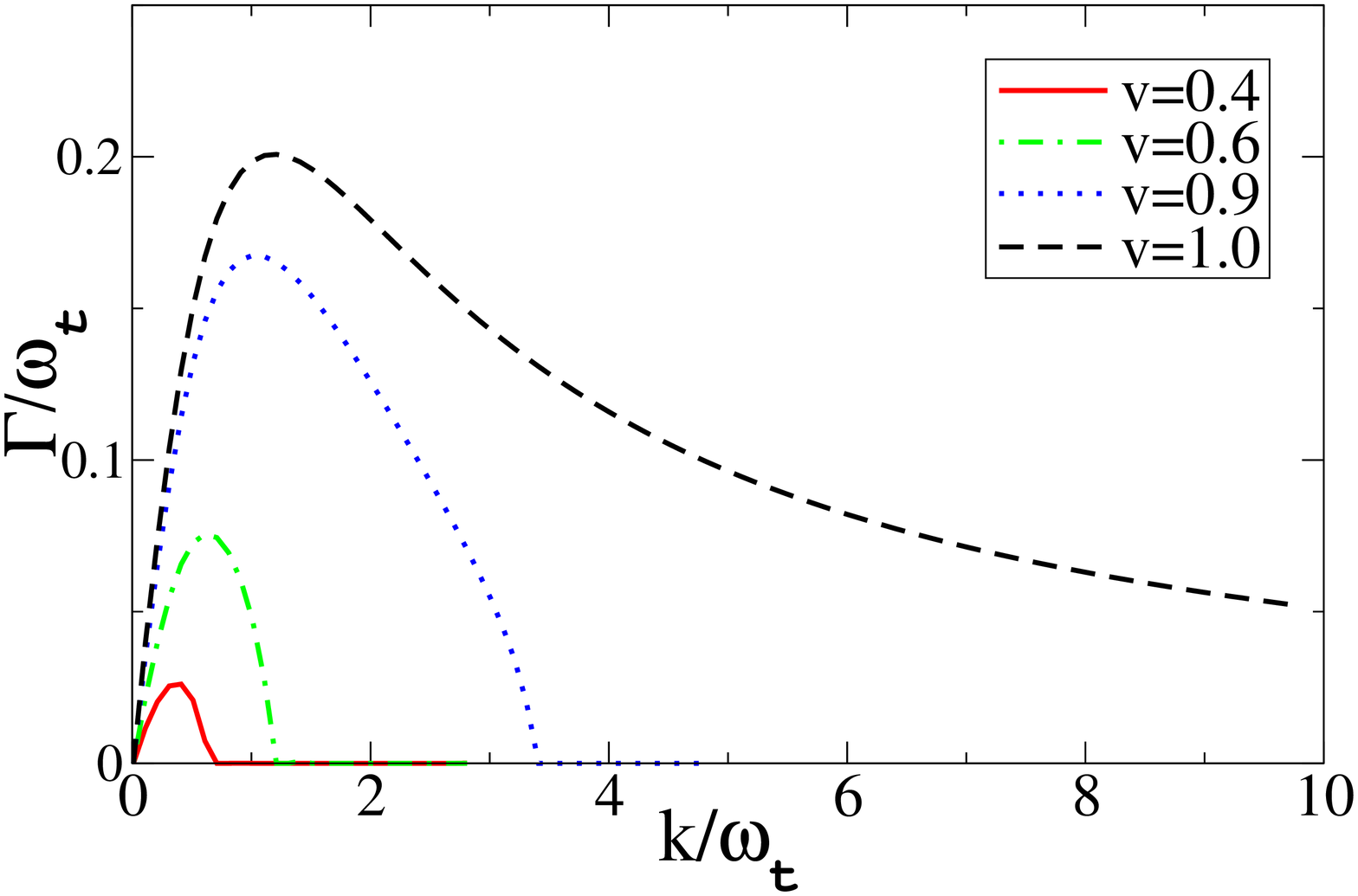}
\includegraphics[width=14pc]{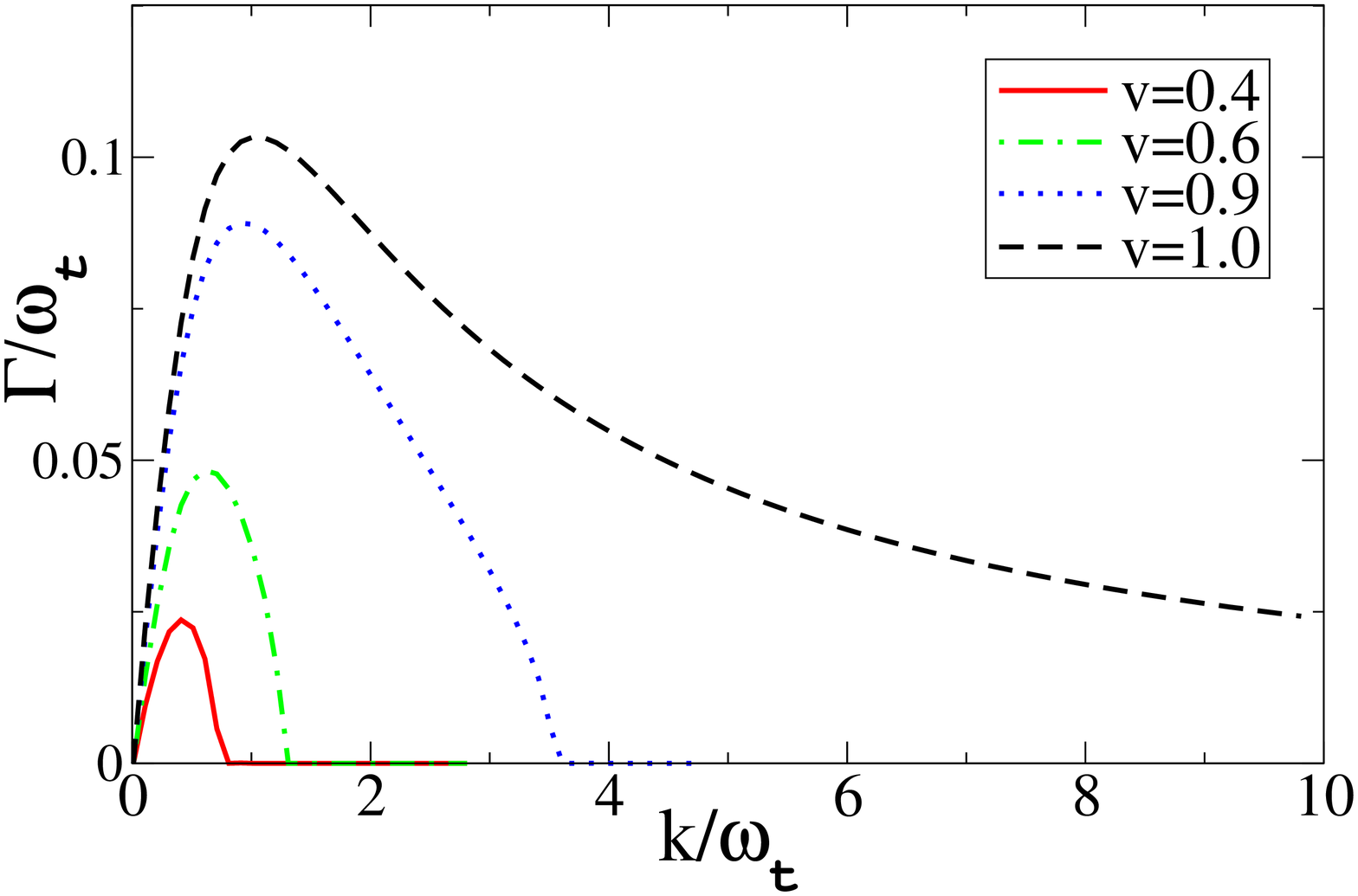}
\caption{\label{perp} Imaginary part of the dispersion law of the unstable  mode  for the system composed by a plasma and a jet with the kinetic theory approach for  $\bf k \perp v$ as a function of the momentum of the mode.  Left panel refers to $b=0.1$   and right panel refers to $b=0.02$. In both case results for four different values of the velocity of the jet,  $|{\bf v}|$, are shown.}  \label{Orthofig1}
\end{center}
\end{figure}

In Figs.~\ref{para} and \ref{perp} we
report the results for the unstable mode obtained with the kinetic theory approach for the case where ${\bf k} \parallel  {\bf v} $ and  ${\bf k} \perp  {\bf v} $, respectively, for different values of the jet velocity,
and different values of $b$.

From our results one concludes  that the most unstable gauge field
mode corresponds to the case ${\bf k} \perp  {\bf v} $. The
largest values of growth rates correspond to high values of jet
velocities. One can also see that the growth rate increases when
increasing the value of $b$.

We can make an estimate for the time scale $t$ for the development of the stream instabilities in the QGP.
In the weak coupling limit, and for   $T \sim 350$ MeV, we find  $t \sim 1-2$ fm/c, meaning that it is
a very fast phenomena.

\section{Energetic considerations}

With the generation of exponentially growing gauge fields it is
clear that there must be an effective transfer of energy from the
jet to the growing chromoelectromagnetic fields. Such a transfer
can be studied by analyzing the energy-momentum $\Theta^{\mu \nu}$
associated to both the partons of the jet and the plasma, as well
as that of the gauge fields.

The energy-momentum tensor due to the quasiparticles  obeys
\be
{\rm Tr} (\delta j_{\mu \,t} F^{\mu \nu}) = \partial_\mu \Theta^{\mu \nu}_{\rm pl} + \partial_\mu \Theta^{\mu \nu}_{\rm jet}
\ee

In collaboration with Michael Strickland \cite{collabor} we are currently analyzing the dynamical evolution
 of the energy
of all the modes in the problem. For such a study we are employing the numerical code of Ref.~\cite{Rebhan:2005re},
originally developed to study Weibel instabilities but now adapted to consider our physical setting.
Our study is performed for a $SU(2)$ non-Abelian group, and  we use a Gaussian distribution to model the jet
distribution function. The parametrization used for this function is such that in a given limit we reproduce
the tsunami-like distribution function. The results of this numerical analysis will be presented elsewhere \cite{collabor}.
They confirm our initial expectations of jet energy loss.

\section{Outlook}

Stream instabilities in the QGP can represent a collective
mechanism for jet quenching. In a weak coupling scenario, this
process is  in principle  much faster than collisional loss, which
is suppressed by an additional factor of $1/g^2$.  Still  the
strength of the effect we are discussing depends not only on $g$,
but also  on  the ratio of densities of the jet and plasma, as
well as in the jet velocity. It is not easy to predict  how the
mechanism we are proposing competes with collisional and radiative
energy loss, so as to assess its relevance for heavy ion
phenomenology. We leave this question for future studies.

Let us stress again that our model of the jet of partons was
extremely crude, but enough for preliminary estimates and also to
study the dynamical evolution of these stream instabilities. One
should localize the jet in space, with the probable associated
effect of having a localized energy deposition. So far we have
only considered the very short time evolution of the  composed
system. At sufficiently long times, collisions should  be
considered, by the addition of the corresponding terms in the
transport equations. Collisions might stop the growth of the
unstable gauge field modes, if those do not saturate due to
non-linear non-Abelian interactions, and they represent a
different sort of mechanism of jet energy loss.
A complete analysis of jet quenching in the QGP should probably include
all these different
energy loss mechanisms.

\section*{Acknowledgements}

C.~M. thanks the  organizers of the YITP Symposium ''Fundamental Problems in Hot and/or Dense QCD"
for the invitation to a very vivid meeting, and to the YITP for hospitality.
This work has been supported by the Ministerio de Educacion y Ciencia (MEC) under Grants No. AYA
2005-08013-C03-02.

\section*{References}


\end{document}